\documentclass{article}

\usepackage{graphicx}

\title{Search for sterile neutrinos as another research objective of ${\theta}_{13}$ 
experiments at reactors}

\author{V. Kopeikin\thanks{kopeykin@polyn.kiae.su}, L. Mikaelyan\thanks{mikaelyan@polyn.kiae.su}, 
V. Sinev\thanks{sinev@polyn.kiae.su}}
\date{Russian Research Centre "Kurchatov Institute", Moscow, Russia}

\begin{document}
\maketitle

Talk given at II Workshop on Future low energy neutrino experiment, Munich, October 9-11, 2003

\begin{abstract}
Searches for mass-3 component in the electron neutrino flavor state $\sin{\theta}_{13}$ and for 
sterile neutrinos can be carried out in the same small mixing angle antineutrino oscillation 
experiment at a reactor. As an example we consider a layout, which involves several movable 
antineutrino spectrometers, stationed at distances 1700 m $-$ 50 m from a reactor. 
The experiment can scan neutrino mass parameter interval $\sim$(0.5$-$0.001) eV$^{2}$ 
and have there typical sensitivity to $\sin^{2}2{\theta}$ at a level of 0.015$-$0.02. 
The signature for sterile neutrino is disappearance observed at mass parameter
${\Delta}m^{2}_{new}$ different from ${\Delta}m^{2}_{atm} \approx 2\times 10^{-3}$ eV$^{2}$. 
In any case existing constraints both on $\sin{\theta}_{13}$ and on sterile neutrinos can 
considerably be improved. 
\end{abstract}

\section*{Introduction} 

Discovery of sterile neutrinos would have a revolutionary impact on neutrino and particle physics.

Sterile neutrinos can hide, mimic or distort reactor antineutrino disappearance pattern in the 
atmospheric oscillation channel.

The notion of sterile neutrinos ${\nu}_s$ was originally introduced by B. Pontecorvo in 1967 y [1]
and later has been considered by many authors: D. Caldwell and R. Mohapatra [2], S. Bilenky, 
C. Giunti and W. Grimus [3], K. Benakli and A. Smirnov [4], B. Kayser [5]. Information on theory 
of sterile (and mirror) neutrinos and references can be found in the recent paper by 
V. Berezinsky, M. Narayan, F. Vissani [6]. 

An experimental hint in favor of sterile neutrinos comes from unconfirmed observations of LSND 
collaboration [7] on ${\nu}_{\mu} \rightarrow {\nu}_e$ transitions. 

An idea how to look for sterile neutrinos at reactors along with $\sin{\theta}_{13}$ was 
proposed in Kurchtov Institute in 1998 y [8]. 

While solar, atmospheric, and laboratory (Super Kamiokande, SNO, KamLAND…) studies are 
understood as only 3-active  neutrino  mixing (see however de Holanda, A. Smirnov, 
hep-ph/0307266) they \underline{do not exclude} some admixture of sterile neutrinos.

\section{How to search for sterile neutrios at reactors}

\underline{In the 3-active neutrino mixing} there are 3 masses and 3 mass parameters:
\begin{eqnarray}
{\Delta}m^{2}_{12} = {\Delta}m^{2}_{solar} \sim (6-8)\times 10^{-5} {\rm eV}^{2}, \nonumber \\
{\Delta}m^{2}_{atm} = {\Delta}m^{2}_{13} \approx {\Delta}m^{2}_{23}\approx 2\times 10^{-3} {\rm eV}^{2} >> {\Delta}m^{2}_{solar}. \nonumber 
\end{eqnarray}

Antineutrino disappearance at distances $L$ = 1000-2000 m from reactor source is governed 
by mass parameter ${\Delta}m^{2}_{atm}$ and by mixing parameter $\sin^{2}2{\theta}_{13}$:
\begin{equation}
P({\nu}_{e} \rightarrow {\nu}_{e}) = 1 - \sin^{2}2{\theta}_{13}\sin^{2}\left(\frac{1.27L{\Delta}m^{2}_{atm}}{E}\right).
\end{equation}

\underline{In 3 active + 3 passive neutrino mixing} there are 6 masses, 15 (!) mass parameters and a 
great number of mixing parameters. 

It can quite happen that at least one of 12 new mass parameters ${\Delta}m^{2}_{new}$ fells 
into the region ${\Delta}m^{2}_{new} \sim (0.5-1.0\times 10^{-3})$ eV$^2$. 

It can be found there in the experiment of Kr2Det type [9] or its modification (some of 
them were discussed in 2002$-$2003 yy in Paris, Alabama and here at TUM), provided the associated 
mixing parameter $\sin^{2}2{\theta}_{s}$ is not too small.

{\bf Antineutrino disappearance, found in a new channel ${\Delta}m^{2}_{new}$ would mean 
existence of sterile neutrino(s).}	

Some part of the $\sin^{2}2{\theta}-{\Delta}m^{2}$ plane is already excluded by the CHOOZ, 
Palo-Verde and Bugey experiments (shaded area in Fig. 1), vast left region is still to be 
explored. 

\section{ Example of layout}

Imagine that a tunnel is built near one 3.2 GW thermal power reactor. We consider five 
identically designed 30 ton target scintillator (movable) detectors, four of them stationed 
in the far position at a distance of 1700 m from the reactor, one 30 ton detector is 
stationed in the near position at 300 m from the reactor. To expand the explored mass 
parameter region towards larger values two small detectors are considered 
at 300 m and 50 m from the reactor.  

Expected neutrino detection rates per 300 days are shown in Table 1

\begin{table}[htb]
\caption{Detector positions, scintillator target masses and $\bar{{\nu}_e}$ detection rates 
per 300 days.}
\vspace{3pt}
\begin{tabular}{c|c|c}
\hline
Distance & Target mass & $\bar{{\nu}_e}$rate/300day \\
  m & ton & \\
\hline
50  & 5 & 1 100 000 \\
300 & 5 &   30 000 \\
300 & 30 &  190 000 \\
1700 & 4 x 30 & 24 000  \\
\hline
\end{tabular} 
\end{table}

\section{Analysis}

Now instead of Eq. (1) we write:

\begin{eqnarray}
P({\nu}_{e} \rightarrow {\nu}_{e}) = 1 - \sin^{2}2{\theta}_{13}\sin^{2}\left(\frac{1.27L{\Delta}m^{2}_{atm}}{E}\right) \nonumber \\
- \sin^{2}2{\theta}_{s} \sin^{2}\left(\frac{1.27L{\Delta}m^{2}_{new}}{E}\right),
\end{eqnarray}
where ${\theta}_s$ and ${\Delta}m^{2}_{new}$ refer to the sterile neutrino. 
We consider two types of data analysis: SHAPE and RATE.
With ONE reactor as ${\bar{\nu}_e}$ source the SHAPE analysis (as we already know) is independent of exact knowledge of:

- Reactor power,	

- Energy spectrum of ${\bar{\nu}_e}$ and its time variations,

- Target volumes and Proton concentrations,

- Absolute efficiencies of ${\bar{\nu}_e}$ detection. 

- Backgrounds can periodically be measured.

The analysis based on comparison of the far/near ${\bar{\nu}_e}$ detection RATES requires good 
knowledge of Ratios of the target volumes and of Antineutrino detection efficiencies.

In both cases NO exact information from the reactor services on reactor power and 
fissile fuel composition is needed for data analysis.

\section{Expected sensitivity}

With 3 years of data taking (300 days/year) most part of the ${\Delta}m^{2}$ range 
(0.5-0.001) eV$^2$ can be searched for ${\theta}_{13}$ and sterile neutrinos with a 
sensitivity of $\sin^{2}2{\theta}_{13}(\sin^{2}2{\theta}_{s}) \sim 0.01-0.015-0.02$ which is in 
general agreement with the analysis performed by P.Huber, M.Lindner, T.Schwetz and W.Winter [10]. 

The limits shown in Fig. 3 were obtained assuming 
energy resolution ${\sigma}_{E} = 0.08 \sqrt{E}$ and the systematics:
${\sigma}_{shape} = 0.5\%, {\sigma}_{rate} = 1\%$.

As can be seen in Fig. 3 the CHOOZ limit on $\sin^{2}2{\theta}_{13}$ at 
${\Delta}m^{2}_{atm} = 2\times 10^{-3}$ eV$^2$ can be improved by a factor 10.

\section{ Instead of discussions}

Question: Should we look for sterile neutrinos in the ${\theta}_{13}$ experiments?

Probability $P$ that we find them is low: $P \sim0.001$

Importance $I$ of finding steriles is very high: $I \sim 1000$

Argument in favor $A$:

$$
A = P \times I \approx 0.001\times 1000 = 1 (!)
$$

Answer: Yes, no doubt, we should look for sterile neutrinos.

\section{Conclusions}

Search for sterile neutrinos at reactors do NOT require much additional effort and can be done 
along with ${\theta}_{13}$

With ONE Reactor and a number of detectors high sensitivity to ${\theta}_{13}$ and steriles can 
be reached.

\section*{Acknowledgments} 

We are grateful to Prof. Yu. Kamyshkov for fruitful discussions of ${\theta}_{13}$ problems. 
We thank Profs. M. Lindner and L. Oberauer for hospitality and beautiful organization of  
Munich workshop. 

\section*{References}
1. B. Pontecorvo, J. Exp. Theor. Phys. 53 1717 (1967). [Sov. Phys. JETP 26 984 (1968)]. \\
2. D. Caldwell and R. Mohapatra, Phys. Rev. D 46, 3259 (1993). \\
3. S. Bilenky, C. Giunti and W. Grimus, Eur. Pys. J. C1, 247 (1998). \\
4. K. Benakli and A. Smirnov, Phys. Rev. Lett. 79, 4314 (1997). \\
5. B. Kayser, hep-ph/9810513. \\
6. V. Berezinsky, M. Narayan, F. Vissani, Nucl. Phys. B 658 (2003) 254. \\
7. LSND Collaboration, Phys. Rev. Lett. 81 (1998) 1774. \\
8. L. Mikaelyan, V. Sinev, Phys. At. Nucl. 62 (1999) 2008, hep-ph/9811228. \\
9. 4. L. Mikaelyan, V. Sinev, Phys. Atom. Nucl. 63 1002 (2000), (hep-ex/9908047); L. Mikaelyan, Nucl. Phys.B (Proc. Suppl.) 87 284 (2000);
hep-ex/9910042; Nucl. Phys. B (Proc. Suppl.) 91 120 (2001), (hep-ex/0008046); V.Martemyanov et al., hep-ex/0211070. \\
10. P.Huber, M.Lindner, T.Schwetz and W.Winter in hep-ph/0303232.

\end{document}